\documentclass[12pt]{article}

\title{\bf Controlling Entanglement Generation in External Quantum Fields}

\author{F. Benatti$^{a,b}$\ and R. Floreanini$^{b,a}$\\  
\small ${}^a$Dipartimento di Fisica Teorica, Universit\`a di Trieste,
Strada Costiera 11,\\ 
\small 34014 Trieste, Italy\\
\small ${}^b$Istituto Nazionale di Fisica Nucleare, Sezione di Trieste,
Strada Costiera 11,\\ 
\small 34014 Trieste, Italy}

\date{\null}

\begin{document}

\maketitle

\begin{abstract}
Two, non-interacting two-level atoms immersed in a common bath can
become mutually entangled when evolving with a Markovian, completely
positive dynamics. For an environment made of external quantum fields,
this phenomenon can be studied in detail: one finds that entanglement 
production can be controlled by varying the bath temperature
and the distance between the atoms. Remarkably, in certain
circumstances, the quantum correlations can persist in the asymptotic 
long-time regime.
\end{abstract}

\section{Introduction}

Independent atoms immersed in external quantum fields 
and weakly coupled to them can be viewed as open systems,
{\it i.e.} as subsystems 
in interaction with an environment \cite{1}-\cite{4}.
The atoms can be usually treated in a non-relativistic approximation,
as independent $n$-level systems, with negligible size, while the environment
is described by a set of quantum fields ({\it e.g.} the electromagnetic field)
in a given quantum state, typically either a temperature state or simply 
the vacuum state. The interaction of the fields with the atoms
is taken to be of dipole type, a well justified approximation
within the weak coupling assumption \cite{5}. 

Even in this simplified setting,
that ignores all intricancies related to the internal atom
structure and the full coupling with the electromagnetic field,
the model is of great relevance both theoretically and
phenomenologically \cite{5}-\cite{7}: indeed, with suitable adaptations,
it is able to capture the main features of the dynamics 
of very different physical systems, like ions in traps,
atoms in optical cavities and fibers, impurities in phonon fields.

Despite this ample range of possible applications and the 
attention devoted to them in the recent literature,
no particular care has often been taken in the derivation
of an acceptable subdynamics for the atoms. 
As a result, time evolutions 
that are not even positive, have been adopted
in order to describe their physical properties.

On the other hand, a mathematically sound and physically consistent time evolution
for the atom subsystem can be obtained using the 
{\it weak coupling limit} procedure \cite{8}: the resulting subdynamics
is described by a one parameter ($\equiv$ time) family of completely
positive maps that form a quantum dynamical semigroup.

In the following we shall outline such a derivation,
and apply the resulting dynamics to the study
of the evolution of a system composed by two, independent atoms.
For simplicity we shall restrict the
attention to two-level atoms in interaction with a collection
of independent, free, massless scalar fields in $3+1$ space-time
dimensions, assumed to be in a state at temperature $T=1/\beta$.

The interaction with an environment
usually leads to decoherence and noise, typical mixing enhancing
phenomena. Therefore, one generally expects that when a bipartite
system is immersed in an environment, quantum correlations
that might have been created before by a direct interaction between the
two subsystems actually disappear. 

However, an external environment can also provide an indirect
interaction between otherwise totally decoupled subsystems
and therefore a mean to correlate them \cite{9}-\cite{14},\cite{6}. 
This phenomenon has first
been established in exactly solvable models \cite{9}: there,
correlations between the two subsystems take place during a
short time transient phase, where the reduced dynamics
of the subsystems contains memory effects.

Remarkably, entanglement generation may also occur in the
Markovian regime, through a purely noisy
mechanism \cite{15,16}. It is precisely this situation that 
is relevant in the analysis of the dynamics of two independent atoms
interacting with the same set of quantum fields.
In the following, we shall study in detail the conditions
that allow the two otherwise indepedent atoms to become
initially entangled through the action of the environment, paying
special attention to the external controllable parameters,
the bath temperature and the spatial distance $\ell$
between the two atoms.
We shall see that for fixed, finite $\ell$, there is always a temperature
below which entanglement generation occurs as soon as
time starts to become nonzero.
Remarkably, it is found that for vanishing $\ell$ the
entanglement thus generated persists even in the long-time 
asymptotic equilibrium state.

\section{Two Atom Master Equation}

We shall deal with a system composed by two, identical
two-level atoms, that start interacting at time $t=\,0$
with a collection of independent, massless, 
scalar quantum fields at temperature $T$.
We are not interested in the details
of the atoms internal dynamics. We shall therefore model them,
in a nonrelativistic way, as simple two-level systems,
which can be fully described in terms of a two-dimensional
Hilbert space.

In absence of any interaction with the external fields,
the single atom internal dynamics will be driven by a $2\times 2$
hamiltonian matrix, that in a given basis can be taken
to assume the most general form:
${\omega\over 2}\, \vec n\cdot\vec\sigma\equiv
{\omega\over 2}\sum_{i=1}^3 n_i\sigma_i\ ,$
where $\sigma_i$, $i=1,2,3$ are the Pauli matrices, $n_i$, $i=1,2,3$
are the components of a unit vector, while $\omega$ represents the
gap between the two energy eigenvalues. 
Then, the atom Hamiltonian $H_S$ is the sum of
two such terms:
\begin{equation}
H_S=H_S^{(1)}+H_S^{(2)}\ ,
\quad
H_S^{(\alpha)}={\omega\over 2}\sum_{i=1}^3 n_i\, \sigma_i^{(\alpha)}
\ ,\quad \alpha=1,2\ ,
\label{1}
\end{equation}
where $\sigma_i^{(1)}=\sigma_i\otimes{\bf 1}$ and
$\sigma_i^{(2)}={\bf 1}\otimes\sigma_i$ are the basis
operators pertaining to the two different atoms.

As mentioned in the Introduction, the interaction of the atoms 
with the external fields is assumed to be weak; it can then
be described by an hamiltonian $H'$ that is linear in both atom and field
variables:
\begin{equation}
H'=\sum_{i=1}^3\Big(\sigma_i^{(1)}\otimes \Phi_i[f^{(1)}]
+\sigma_i^{(2)}\otimes \Phi_i[f^{(2)}]\Big)\ .
\label{2}
\end{equation}
The operators
$\Phi_i(t,\vec x)$ represent the external quantum fields,
taken to be spinless and massless for simplicity. 
They evolve in time as free relativistic fields with a standard
Hamiltonian $H_\Phi$.
The atoms are assumed to have a spatial extension described by the two
functions $f^{(\alpha)}(\vec x)$, $\alpha=1,2$, taken to have
a common profile $f(\vec x)$. To be more specific, we shall choose 
for the atoms a spherically symmetric shape of infinitesimal size $\varepsilon$:
\begin{equation}
f(\vec x)={1\over\pi^2}{(\varepsilon/2)\over \big[|\vec x|^2 
+(\varepsilon/2)^2\big]^2}\ .
\label{3}
\end{equation}
Further, without loss of generality,
the first atom can be positioned at the origin of the reference frame,
so that one can assume $f^{(1)}(\vec x)\equiv f(\vec x)$, while the second is
displaced by an amount $\vec\ell$ with respect to it, and therefore: 
$f^{(2)}(\vec x)= f(\vec x+\vec\ell\,)$. Since the atom-field interaction 
takes place on the whole region occupied by the atoms, 
the field operators entering the interaction Hamiltonian 
above are smeared over the atom size:
\begin{equation}
\Phi_i[f^{(\alpha)}]=\int d^3 x\, f^{(\alpha)}(\vec x\,)\, 
\Phi_i(0,\vec x\,)\ ,\quad \alpha=1,2\ .
\label{4}
\end{equation}

The total Hamiltonian $H$ describing the complete system, the two atoms
together with the external fields $\Phi_i$, can thus be written as
\begin{equation}
H=H_S +H_\Phi +\lambda H'\equiv H_0 + \lambda H'\ ,
\label{5}
\end{equation}
with $\lambda$ a small coupling constant.
It generates the evolution in $t$ of the corresponding 
total density matrix $\rho_{\rm tot}$,
$\partial_t\rho_{\rm tot}(t)=-i[H,\ \rho_{\rm tot}(t)]$,
starting at $t=\,0$ from the initial
configuration: $\rho_{\rm tot}(0)$;
we shall assume the atom and the fields to be initially prepared 
in an uncorrelated state, with the fields in
the temperature state $\rho_\beta$ and the atoms in a
generic initial state $\rho(0)$, so that
$\rho_{\rm tot}(0)=\rho(0)\otimes \rho_\beta$.

At this point, being interested in studying the dynamics of the two atoms,
one conveniently integrates over the unobserved field degrees of freedom 
and concentrate on the analysis of the reduced time evolution,
formally given by the transformation map:
$\rho(0)\mapsto\rho(t)\equiv{\rm Tr}_\Phi[\rho_{\rm tot}(t)]$.

The derivation of a physically consistent master equation
for the reduced density matrix 
$\rho(t)\equiv{\rm Tr}_\Phi[\rho_{\rm tot}(t)]$ is notoriously
tricky, requiring an {\it a priori}
unambigous separation between subsytem and environment, 
besides a sufficiently weak interaction between the two \cite{1}-\cite{3}.
Generally speaking, this distinction can be achieved when the
correlations in the environment decay much faster than the
characteristic evolution time of the subsystem
alone, given by the inverse of its typical energy scale.
In such a case, in the limit of weak couplings, the changes in the
evolution of the subsystem occur on time scales that are very long, so large
that the details of the internal environment dynamics result irrelevant.
This is precisely the situation that occurs for the system under study:
indeed, in typical instances, 
the differences between the atomic internal energy levels result much smaller
than the field correlation decay constants so that a clear distinction between
subsystem and environment is authomatically achieved.

In practice, the dynamics of the reduced system is obtained
by suitably rescaling the time variable, $t\to t/\lambda^2$
and then taking the limit $\lambda\to 0$, following the 
mathematically precise procedure of the {\it weak coupling limit}
\cite{8},\cite{1}-\cite{3}.
The reduced density matrix $\rho(t)$ is then found to obey 
the following evolution equation:
\begin{equation}
{\partial\rho(t)\over \partial t}= L_{H_S}[\rho(t)]
 + {\cal D}^\sharp[\rho(t)]\ ,\qquad L_{H_S}[\rho]\equiv-i \big[H_S,\, \rho\big]\ ,
\label{6}
\end{equation}
where
\begin{equation}
{\cal D}^\sharp[\,\cdot\,]=-\lim_{T\to\infty}{1\over T}\int_0^T ds\
{\cal U}(-s)\ {\cal D}\ {\cal U}(s)\,[\, \cdot\,]\ ,\qquad
{\cal U}(s)=e^{sL_{H_S}}\ ,\label{7}
\label{7a}
\end{equation}
and
\begin{equation}
{\cal D}[\rho]=\int_0^\infty dt\ {\rm Tr}\Big( \big[e^{iH_0 t}\, H'\, e^{-iH_0 t},\big[H',\, 
\rho\otimes\rho_\beta\big]\big]\Big)\ .
\label{8}
\end{equation}
For the case at hand, the integrals in (\ref{7}) and (\ref{8}) can be explicitly
computed and the master equation for $\rho(t)$ written down without any ambiguity.
It takes a Kossakowski-Lindblad form \cite{17,18}
\begin{equation}
{\partial\rho(t)\over \partial t}= -i \big[H_{\rm eff},\, \rho(t)\big]
 + {\cal L}[\rho(t)]\ ,
\label{9}
\end{equation}
with
\begin{equation}
H_{\rm eff}=H_S-\frac{i}{2}\sum_{\alpha,\beta=1}^2\sum_{i,j=1}^3
H_{ij}^{(\alpha\beta)}\ \sigma_i^{(\alpha)}\,\sigma_j^{(\beta)}\ ,
\label{10}
\end{equation}
and
\begin{equation}
{\cal L}[\rho]={1\over2} \sum_{\alpha,\beta=1}^2\sum_{i,j=1}^3 C_{ij}^{(\alpha\beta)}\big[2\, 
\sigma_j^{(\beta)}\rho\,\sigma_i^{(\alpha)} 
-\sigma_i^{(\alpha)}\sigma_j^{(\beta)}\, \rho 
-\rho\,\sigma_i^{(\alpha)}\sigma_j^{(\beta)}\big]\ .
\label{11}
\end{equation}
The coefficients of the Kossakowski matrix $C_{ij}^{(\alpha\beta)}$
and of the effective Hamiltonian $H_{\rm eff}$ are determined by
the field correlation functions in the thermal state $\rho_\beta$:
\begin{equation}
G_{ij}^{(\alpha\beta)}(t-t')=\int d^3x\, d^3y\, f^{(\alpha)}(\vec x)\,
f^{(\beta)}(\vec y)\ \langle \Phi_i(t,\vec x)\Phi_j(t',\vec y)\rangle\ ,
\label{12}
\end{equation}
through their Fourier,
\begin{equation}
{\cal G}_{ij}^{(\alpha\beta)}(z)=\int_{-\infty}^{\infty} dt \, e^{i z t}\, 
G_{ij}^{(\alpha\beta)}(t)\ ,
\label{13}
\end{equation}
and Hilbert transform,
\begin{equation}
{\cal K}_{ij}^{(\alpha\beta)}(z)=\int_{-\infty}^{\infty} dt \, {\rm sign}(t)\, 
e^{i z t}\, G_{ij}^{(\alpha\beta)}(t)=
\frac{P}{\pi i}\int_{-\infty}^{\infty} dw\ \frac{ {\cal G}_{ij}^{(\alpha\beta)}(w) }{w-z}
\ ,
\label{14}
\end{equation}
respectively ($P$ indicates principle value). More specifically, one finds:
\begin{equation}
C_{ij}^{(\alpha\beta)}=\sum_{\xi=+,-,0} \sum_{k,l=1}^3
{\cal G}_{kl}^{(\alpha\beta)}(\xi\omega)\, 
\psi_{ki}^{(\xi)}\, \psi_{lj}^{(-\xi)}\ ,
\label{15}
\end{equation}
%
%
and similarly for $H_{ij}^{(\alpha\beta)}$, with
${\cal G}_{kl}^{(\alpha\beta)}(\xi\omega)$ replaced by
${\cal K}_{kl}^{(\alpha\beta)}(\xi\omega)$, where
\begin{equation}
\psi_{ij}^{(0)}=n_i\, n_j\ ,\qquad 
\psi_{ij}^{(\pm)}={1\over 2}\big(\delta_{ij} - n_i\, n_j\pm i\epsilon_{ijk} n_k\big)\ .
\label{16}
\end{equation}
are the components of auxiliary three-dimensional tensors.%
\footnote{We omit the details of the derivation and refer to the Appendix
of \cite{16} for an outline of the needed techniques.}
Being the sum of three positive terms,
the matrix $C_{ij}^{(\alpha\beta)}$ turns out to be positive, so that the 
dynamical semigroup generated by (\ref{9}) is composed by
completely positive maps: this is the result of adopting
a mathematically well-defined formalism \cite{8}.
On the other hand, let us remark
that direct use of the standard second order perturbative approximation
({\it e.g.} see \cite{6,7}) often leads to physically inconsistent
results, giving a finite time evolution for $\rho(t)$ that in general does 
not preserve the positivity of probabilities.

The expressions in (\ref{13}) and (\ref{14}) can be explicitly computed by
noting that the fields are taken to be independent and assumed to obey a free evolution,
so that:
\begin{equation}
\langle \Phi_i(x)\Phi_j(y)\rangle\equiv {\rm Tr}\big[\Phi_i(x)\Phi_j(y)\rho_\beta\big]
=\delta_{ij}\, G(x-y)\ ,
\label{17}
\end{equation}
where $G(x-y)$ is the standard four-dimensional Wightmann function for
a single relativistic scalar field in a state at inverse temperature $\beta$,
formally given by:
\begin{equation}
G(x)=\int \frac{d^4 k}{(2\pi)^{3}}\, \theta(k^0)\, \delta(k^2)
\Big[\big(1+{\cal N}(k^0)\big)\, e^{-ik\cdot x}+{\cal N}(k^0)\, e^{ik\cdot x}
\Big]\ ,
\label{18}
\end{equation}
where
\begin{equation}
{\cal N}(k^0)=\frac{1}{e^{\beta k^0} -1}\ .
\label{19}
\end{equation}
Note that, as a result, the correlations in (\ref{12}) involve 
${\hat f}(\vec k)=\int d^3x\, e^{i\vec k\cdot\vec x}\, f(\vec x)$,
namely the Fourier transform
of the shape function $f(\vec x)$ in (\ref{3}); it can be easily computed to be
${\hat f}(\vec k\,)=e^{-|\vec k\,| \varepsilon/2}$.
Being a function of the modulus $|\vec k\,|$ only,
this contribution can be conveniently attached to the
definition of the Wightmann function $G(x)$, so that the integrand 
in (\ref{18})
gets an extra $e^{-\varepsilon k^0}$ overall factor. This damping term
assures now the convergence of the integral (\ref{18}) and corresponds
to the usual $i\varepsilon$ prescription for the Wightmann function;
in the present setting, it arise as a remnant of the (infinitesimal)
size of the atoms.

The behaviour of the field correlations in (\ref{12}) is also crucial
for assuring the convergence of the evolution equation of the reduced density
matrix $\rho(t)$ to the limit (\ref{6});
indeed, one shows \cite{8} that such limit exists only when the
combination $|G_{ij}^{(\alpha\beta)}(t)|(1+t)^\eta$ is integrable 
on the positive half real line, for some $\eta>0$.
In the case of massless fields considered here, this condition
is assured by the $1/t^2$ fall off at infinity of the
Wightmann function.

Using (\ref{17}) and (\ref{18}), the Fourier transform in (\ref{13})
can now be explicitly evaluated (in the limit of vanishing $\varepsilon$):
\begin{equation}
{\cal G}_{ij}^{(\alpha\beta)}(z)=
\delta_{ij}\ {\cal G}^{(\alpha\beta)}(z)\ ,
\label{20}
\end{equation}
with:
\begin{eqnarray}
\nonumber
&&{\cal G}^{(11)}(z)={\cal G}^{(22)}(z)=\frac{1}{2\pi} \frac{z}{1-e^{-\beta z}}\ ,\\
\label{21}
&&{\cal G}^{(12)}(z)={\cal G}^{(21)}(z)=\frac{1}{2\pi} \frac{z}{1-e^{-\beta z}}\ 
\frac{\sin(\ell z)}{\ell z}\ ,
\end{eqnarray}
where $\ell$ denotes the modulus of the displacement vector $\vec\ell$;
then, recalling (\ref{14}), for the Hilbert transform one similarly
finds:
\begin{equation}
{\cal K}_{ij}^{(\alpha\beta)}(z)=
\delta_{ij}\ {\cal K}^{(\alpha\beta)}(z)\ ,\qquad
{\cal K}^{(\alpha\beta)}(z)=\frac{P}{\pi i}\int_{-\infty}^{\infty} dw\ 
\frac{ {\cal G}^{(\alpha\beta)}(w) }{w-z}\ .
\label{22}
\end{equation}
With these results, The Kossakowski matrix can finally be written as:
\begin{eqnarray}
\nonumber
&&C^{(11)}_{ij}=C^{(22)}_{ij}=A\, \delta_{ij}-iB\, \epsilon_{ijk}\, n_k + C\, n_i\, n_j\ ,\\
\label{23}
&&C^{(12)}_{ij}=C^{(21)}_{ij}=A'\, \delta_{ij}-iB'\, \epsilon_{ijk}\, n_k + C'\, n_i\, n_j\ ,
\end{eqnarray}
where the quantities $A$, $B$, $C$, $A'$, $B'$ and $C'$ depend on the system frequency $\omega$,
the inverse temperature $\beta$ and the separation $\ell$ between the two atoms:
\begin{eqnarray}
\label{24}
A&=&{\omega\over 4\pi}
\bigg[{1+e^{-\beta\omega}\over 1-e^{-\beta\omega}}\bigg]\ ,\hskip 2cm 
A'={\omega\over 4\pi}
\bigg[{1+e^{-\beta\omega}\over 1-e^{-\beta\omega}}\bigg]\ 
\frac{\sin(\omega\ell)}{\omega\ell}\ ,\\
\label{25}
B&=&{\omega\over 4\pi}\ ,\hskip 4cm
B'={\omega\over 4\pi}\ \frac{\sin(\omega\ell)}{\omega\ell}\ ,\\
\label{26}
C&=&{\omega\over 4\pi}
\bigg[{2\over\beta\omega}-{1+e^{-\beta\omega}\over 1-e^{-\beta\omega}}\bigg]\ ,
\hskip 1cm
C'={\omega\over 4\pi}
\bigg[{2\over\beta\omega}-{1+e^{-\beta\omega}\over 1-e^{-\beta\omega}}\
\frac{\sin(\omega\ell)}{\omega\ell}\bigg]\ .
\end{eqnarray}

On the other hand, the effective Hamiltonian $H_{\rm eff}$ naturally splits
into two contributions: $H_{\rm eff}={\tilde H}_S+H_{\rm eff}^{(12)}$.
The first one has the same form of the starting system Hamiltonian
in (\ref{1}), but with a redefined frequency:
\begin{equation}
\tilde\omega=\omega+i\big[ {\cal K}^{(11)}(-\omega) - {\cal K}^{(11)}(\omega)\big]\ ,
\label{27}
\end{equation}
while the second one corresponds to an environment generated direct coupling
among the two atoms:
\begin{eqnarray}
\nonumber
H_{\rm eff}^{(12)}=-\frac{i}{2}\sum_{i,j=1}^3 && \Big(
\big[ {\cal K}^{(12)}(-\omega) + {\cal K}^{(12)}(\omega)\big]\ \delta_{ij}\\
&&+\big[ 2{\cal K}^{(12)}(0)-{\cal K}^{(12)}(-\omega) - {\cal K}^{(12)}(\omega)\big]
\, n_i n_j\Big)\ \sigma_i\otimes\sigma_j\ .
\label{28}
\end{eqnarray}
These results for the Hamiltonian contributions 
require some further comments. 
Recalling (\ref{21}), the definition of ${\cal K}^{(11)}(z)$
in (\ref{22}) can be split as (similar results hold also for
${\cal K}^{(12)}(z)$):
\begin{eqnarray}
\label{29}
{\cal K}^{(11)}(z)&=&
\frac{1}{2\pi^2 i}\Bigg[ P
\int_0^\infty dw\
\frac{w}{w-z}\\
\label{30}
&&\hskip 2.5cm +P\int_0^\infty dw\ \frac{w}{1-e^{\beta w}}
\Bigg(\frac{1}{w+z}-\frac{1}{w-z}\Bigg)\Bigg]\ ,
\end{eqnarray}
into a vacuum and a temperature-dependent piece.
Although not expressible in terms of elementary functions, the temperature
dependent second term is a finite, odd function of $z$, vanishing
as $\beta$ becomes large, {\it i.e.} in the limit of zero temperature
(and as such, it does not contribute to $H_{\rm eff}^{(12)}$ in (\ref{28})).
The first contribution in (\ref{29}) is however divergent.
As a consequence, despite some cancellations 
that occur in (\ref{27}) and (\ref{28}), the effective Hamiltonian
$H_{\rm eff}$ turns
out in general to be infinite, and its definition requires the introduction 
of a suitable cutoff and a renormalization procedure.

This is a well known fact, and has nothing to do with the
weak-coupling assumptions used in deriving the master equation.
Rather, the appearance of the
divergences is due to the non-relativistic treatment of
the two-level atoms, while any sensible calculation
of energy shifts would have required the use 
of quantum field theory techniques \cite{5}.

In our quantum mechanical setting, the procedure needed
to make $H_{\rm eff}$ well defined is therefore
clear: perform a suitable
temperature independent subtraction, so that
the expressions in (\ref{27}) and (\ref{28}) reproduce the correct
quantum field theory result, obtained by considering
the fields in the vacuum state.

In the following, we shall be interested in analyzing
the temperature dependent effects described by the
master equation (\ref{9}); all standard, vacuum generated
Hamiltonian contributions will be therefore ignored.

\section{Environment Induced Entanglement Generation}

Using the explicit form for the master equation
derived in the previous Section, one can now
investigate whether the thermal bath made of
free fields can actually entangle the two indepedent atoms.
Since we are dealing with a couple of two-level systems,
this can be achieved with the help of the 
partial transposition criterion \cite{19,20}:
a two-atom state $\rho(t)$ results entangled at time $t$
if and only if the operation of partial transposition
does not preserve its positivity.

We shall first consider the possibility of entanglement creation
at the beginning of the evolution. 
Without loss of generality, one can limit the considerations
to pure, separable initial states, and therefore take: 
\begin{equation}
\label{31}
\rho(0)=\vert \varphi\rangle\langle \varphi\vert\otimes 
\vert \psi\rangle \langle \psi\vert\ ;
\end{equation}
indeed, if the environment is unable to create 
entanglement out of pure states, 
it will certainly not entangle their mixtures.
Then, let us examine the behavior in a neighborhood of
$t=\,0$ of the quantity
\begin{equation}
{\cal Q}(t)=\langle\chi\vert\, \tilde{\rho}(t)\, \vert\chi\rangle\ ,
\label{32}
\end{equation}
where the tilde 
signifies partial transposition, {\it e.g.} with respect to the second factor,
and $|\chi\rangle$ is any $4$-dimensional vector. 
The two atoms, initially prepared in a 
state $\rho(0)\equiv\tilde\rho(0)$ as in (\ref{31}), will surely
become entangled
if there exists a suitable vector $|\chi\rangle$, such that: 
\hbox{{\sl i)} ${\cal Q}(0)=\,0$} and 
\hbox{{\sl ii)} $\partial_t {\cal Q}(0)<0$}.
In fact, when $\partial_t{\cal Q}(0)>0$ for all choices 
of the initial state $\rho(0)$ and probe vector $|\chi\rangle$, 
entanglement can not be generated by the environment, since $\tilde\rho$
remains positive. Clearly, the vector $\vert\chi\rangle$ need be
chosen entangled,
since otherwise ${\cal Q}$ is never negative.

Note that study of the behavior of the quantity ${\cal Q}(t)$ near $t=\,0$
allows a very explicit analysis of entanglement generation.
Indeed, 
$\partial_t {\cal Q}(0)$ can be easily computed through the 
time derivative $\partial_t\tilde\rho(0)$, that in turn can be obtained
by taking the partial transposition
of the r.h.s. of (\ref{9}) (with $H_{\rm eff}$ set to zero, as explained above).
In this way, one can construct a test of entanglement creation, valid
for any probe vector $|\chi\rangle$.

In order to show this,
consider first the orthonormal basis 
$\{|\varphi\rangle,\ |\tilde\varphi\rangle\}$,
$\{|\psi\rangle,\ |\tilde\psi\rangle\}$, obtained by augmenting
with the two states $|\tilde\varphi\rangle$ and
$|\tilde\psi\rangle$ the ones that define
$\rho(0)$ in (\ref{31}). 
They can be both unitarily rotated to the standard basis
$\{|-\rangle,\ |+\rangle\}$ of $\sigma_3$:
\begin{eqnarray}
&&|\varphi\rangle= U |-\rangle \qquad |\tilde\varphi\rangle= U |+\rangle\ ,
\nonumber\\
&&|\psi\rangle= V |-\rangle \qquad |\tilde\psi\rangle= V |+\rangle\ .
\label{33}
\end{eqnarray}
Similarly, the unitary transformations $U$ and $V$ induce orthogonal 
transformations
$\cal U$ and $\cal V$, respectively, on the Pauli matrices:
\begin{equation}
U^\dagger \sigma_i U=\sum_{j=1}^3 {\cal U}_{ij}\sigma_j\ ,\quad
V^\dagger \sigma_i V=\sum_{j=1}^3 {\cal V}_{ij}\sigma_j\ .
\label{34}
\end{equation}
Direct computation then shows that $\partial_t{\cal Q}(0)$
can be written as a quadratic form in the independent components
of the probe vector $|\chi\rangle$, with coefficients
that involve the four
$3\times3$ matrices $C^{(11)}$, $C^{(22)}$, $C^{(12)}$, $C^{(21)}$
that form the Kossakowski coefficients given in (\ref{15}).
As a consequence, vectors $|\chi\rangle$ exist making this form
negative, {\it i.e.} $\partial_t{\cal Q}(0)<0$,
if and only if its corresponding discriminant is negative;
explicitly:
\begin{equation}
\langle u | C^{(11)} | u \rangle \, \langle v | \big({C^{(22)}}\big)^T
| v 
\rangle <
\big|\langle u | {\cal R}e\big(C^{(12)}\big)| v \rangle \big|^2\ ,
\label{35}
\end{equation}
where $T$ means matrix transposition.
The three-dimensional vectors $|u\rangle$ and $|v\rangle$ 
contain the information about the starting factorized state (\ref{31}):
their components can be in fact expressed as:
\begin{equation}
u_i=\sum_{j=1}^3 {\cal U}_{ij}\, \langle +| \sigma_j |-\rangle\ ,\quad
v_i=\sum_{j=1}^3 {\cal V}_{ij}\, \langle -| \sigma_j |+\rangle\ .
\label{36}
\end{equation}
Therefore, the external quantum fields will be able to entangle the two 
atoms evolving with the Markovian dynamics generated by (\ref{9}) 
and characterized by
the Kossakowski matrix (\ref{15}), if there exists an initial
state $|\varphi\rangle\langle \varphi|\otimes |\psi\rangle\langle \psi|$,
or equivalently orthogonal transformations $\cal U$ and $\cal V$,
for which the inequality (\ref{35}) is satisfied.

That this is indeed the case for the matrices $C^{(\alpha\beta)}$
in (\ref{23}) can be easily shown. First note that, without loss of generality,
the unit vector $\vec n$ that defines the
internal atom Hamiltonian in (\ref{1}) can be 
oriented along the third axis. 
Consider then the initial state
$\rho(0)=|-\rangle\langle -|\otimes |+\rangle\langle +|$,
constructed out of the eigenstates of the single atom Hamiltonian.
Recalling the definitions (\ref{31}) and (\ref{33}), 
one then finds $U={\bf 1}$,
so that the three-dimensional vector $|u\rangle$
has components $u_i=\{1, -i, 0\}$, and further $v_i=u_i$.
Then, the inequality (\ref{35}) reduces to:
\begin{equation}
R^2 + S^2>1\ ,\qquad R\equiv{B\over A}={1-e^{-\beta\omega}
\over 1+e^{-\beta\omega}}\ ,
\qquad S=\frac{\sin(\omega\ell)}{\omega\ell}\ .
\label{37}
\end{equation}
Although both $R$ and $S$ take values in the interval $[0,\, 1]$, 
one can easily
make the sum of their squares to exceed unity by adjusting
the inverse temperature $\beta$ and the atom separation $\ell$. In particular,
for a given, finite separation $\ell$, one can always find a temperature
below which the inequality in (\ref{37}) is satisfied and therefore
entanglement created. The case of
vanishing separation is even more striking, since
the inequality (\ref{37}) reduce to $R>0$, which is always
satisfied, except in the limit of an infinite bath temperature.

Analogous results hold also in the case of a zero temperature bath.
Indeed, in this case $R=1$, and entanglement is generated for any finite
separation of the two atoms (similar conclusions have also been
reported before in Ref.\cite{21}).

The role played by the two considered control parameters, the bath
temperature and the atom separation, in triggering entanglement
creation is now apparent. The temperature of the external
environment determines the amount of noise that is induced in
the dynamics of the two independent atoms. Noise is known
to reduce quantum correlations,
and indeed, the higher the temperature,
the less effective is the entanglement power of the bath.
Environment induced entanglement generation is nevertheless
a robust phenomenon: it always occurs except in the limit
of an infinitely large temperature.

A similar role is played by the second control parameter,
the spacial atom separation: entanglement enhancement
is more effective the closer the two atoms are, and turns
out to be impossible only for an infinitely large separation.
The interplay between the effects of these two control
parameters is neatly summarized by the inequality (\ref{37}).

To our knowledge, this is the first instance of open quantum
system control through the bath parameters, and not via
the subsystem Hamiltonian \cite{22,23}.
This approach, which might prove very fruitful in the
field of quantum information, is still in a very preliminary
stage; further developments are presently under study
and will be reported elsewhere.

Finally, note that the above choice for $\rho(0)$ is not restrictive:
one can always use the transformations in (\ref{33}) to map 
it to the generic initial state (\ref{31}), at the expences of using
suitably rotated $|u\rangle$ and $|v\rangle$, as given in (\ref{36}).
As a result, the expression of the entanglement test in (\ref{35})
becomes more complicated than in (\ref{37}), but the final conclusions
remain unchanged.

\section{Asymptotic Entanglement}

On the basis of the anlysis presented in the previous Section
two atoms, initially prepared in a separate state,
will generically become entangled as a result of their 
independent interaction with a bath made of thermal quantum fields:
quantum correlations are generated among the two
atoms as soon as $t>0$. The test
in (\ref{35}), on which this conclusion is based, is however
unable to determine the fate of this
quantum correlations, as time becomes large.

On general grounds, one expects that the effects 
of decoherence and dissipation that
counteract entanglement production
be dominant at large times, so that 
no entanglement is left in the end.
This is precisely the conclusion that one obtains
by a careful analysis of the structure of the dynamics generated
by the master equation in (\ref{9}), with Kossakowski
coefficients as in (\ref{23})-(\ref{26}): the
asymptotic equilibrium state for the two atoms
turns out to be always separate for $\ell$ finite
(a detailed account of this result is beyond the scope of the
present work, and will be reported elsewhere).

However, the case of a vanishing atom separation is again
special and deserves a closer examination. Note that in such a situation, the
matrices in (\ref{23}) become all equal 
$C^{(11)}=C^{(22)}=C^{(12)}=C^{(21)}$.
This particular choice for the Kossakowski matrix is also adopted in
the description of the phenomenon of resonance fluorescence
\cite{24,7}. 
Therefore,  the discussion of the vanishing $\ell$ limit is of
relevance also from the phenomenological point of view.

The presence of an equilibrium state $\rho^\infty$ can be in general determined
by setting to zero the r.h.s. of the evolution equation (\ref{9}).
As previously explained, we shall ignore the Hamiltonian piece
since it can not give rise to temperature dependent entanglement phenomena,
and concentrate on the study of the effects induced by the
dissipative part; the equilibrium condition reduces then to
${\cal L}[\rho^\infty]=\,0$.
Direct computation leads to the following result:
\begin{equation}
\rho^\infty={1\over4 }\bigg[{\bf 1}\otimes{\bf 1}-
a\sum_{i=1}^3 n_i\big({\bf 1}\otimes\sigma_i +
\sigma_i\otimes{\bf 1}\big)
+ \sum_{i,j=1}^3 \big(b\, \delta_{ij}+c\, n_i n_j\big)\ \sigma_i\otimes\sigma_j\bigg]\ ,
\label{38}
\end{equation}
with
\begin{equation}
a=\frac{R}{3+R^2}(\tau+3)\ ,\qquad b=\frac{\tau-R^2}{3+R^2}(\tau+3)\ ,\qquad
c=R\, a\ .
\label{39}
\end{equation}
Here, $R=B/A$ is the temperature dependent ratio already introduced in (\ref{37}),
while the parameter $\tau=\frac{1}{4} \sum_{i=1}^3{\rm Tr}[\rho(0)\, (\sigma_i\otimes\sigma_i)]$
contains the dependence on the initial state (positivity of $\rho(0)$ requires
$-3\leq\tau\leq1$): the equilibrium state is therefore
not unique. In spite of this, remarkably, the asymptotic state (\ref{38}) 
turns out to be still entangled.

To explicitly show this, one can as before act with the operation of
partial transposition on $\rho^\infty$ to see whether negative eigenvalues
are present. Alternatively, one can resort to one of the available
entanglement measures and concurrence appears here to be 
the more appropriate: its value ${\cal C}[\rho]$ ranges from zero, for 
separable states, to one, for fully entangled states, like the Bell
states \cite{25}-\cite{27}.
In the case of the state $\rho^\infty$ above, 
one finds
\begin{equation}
{\cal C}[\rho^\infty]={\rm max}\Bigg\{{\big(3-R^2\big)\over2\big(3+R^2\big)}\,
\bigg[ {5R^2-3\over 3-R^2} -\tau\bigg],\ 0\Bigg\}\ .
\label{40}
\end{equation}
This expression is indeed nonvanishing, provided we start with
an initial state $\rho(0)$ for which
\begin{equation}
\tau< {5R^2-3\over 3-R^2}\ .
\label{41}
\end{equation}
The concurrence is therefore a linearly decreasing function of $\tau$,
starting from its maximum  ${\cal C}[\rho^\infty]=1$ for $\tau=-3$ 
and reaching zero at $\tau=(5R^2-3)/(3-R^2)$.

This result is remarkable, since it implies that the dynamics
in (\ref{9}) not only can initially generate entanglement: 
it can continue to enhance it even in the asymptotic long time
regime.%
\footnote{Similar results hold in the case of uniformily accelerating
atoms \cite{16}; in that case, entanglement generation is controlled
by the value of the proper acceleration.}
In other terms, put the two atoms in the same place and
prepare them at $t=\,0$ in a separable 
state; then, provided the condition (\ref{41}) is satisfied,
their long time equilibrium state will turn out to be entangled.

The simplest example of a separable state is provided
by the direct product of pure states as in (\ref{31}).
When $|\varphi\rangle$ and $|\psi\rangle$ are orthogonal,
so that $\tau=-1$, one explicitly finds:
${\cal C}[\rho^\infty]={\rm max}\big\{2R^2/ (3+R^2),\, 0\big\}$. 
Notice that ${\cal C}[\rho^\infty]$ 
reaches its maximum value of $1/2$ when $R=1$, {\it i.e.} at zero temperature,
while it vanishes when the temperature
becomes infinitely large, {\it i.e.} $R=\,0$. As already remarked, this has to
be expected, since in this case the decoherence effects
of the bath become dominant.

\end{document}